\documentclass[aip,jmp,amsmath,amssymb,reprint,]{revtex4-1}
\pdfoutput=1
\usepackage{graphicx}% Include figure files
\usepackage{dcolumn}% Align Table columns on decimal point
\usepackage{bm}% bold math
\usepackage{epstopdf}
\usepackage{amssymb}
\usepackage{color}
\usepackage[colorlinks=true, letterpaper=true, pdfstartview=FitV, linkcolor=blue, citecolor=blue, urlcolor=blue]{hyperref}
\usepackage{lineno}
\begin{document}
\title{First-principles study the structural, magnetic, optical properties and doping effect in chromium arsenide}

\author{Na Kang}
\author{Wenhui Wan}
\email{wwh@ysu.edu.cn}

\author{Bu-Sheng Wang}
\affiliation{State Key Laboratory of Metastable Materials Science and Technology $\&$ Key Laboratory for Microstructural Material Physics of Hebei Province, School of Science, Yanshan University, Qinhuangdao, 066004, People's Republic}

\author{Kai-Cheng Zhang}
\affiliation{College of Mathematics and Physics, Bohai University, Jinzhou 121013, China}

\author{Yong Liu}
\email{yongliu@ysu.edu.cn or ycliu@ysu.edu.cn}
\affiliation{State Key Laboratory of Metastable Materials Science and Technology $\&$ Key Laboratory for Microstructural Material Physics of Hebei Province, School of Science, Yanshan University, Qinhuangdao, 066004, People's Republic}

\begin{abstract}
We systematically study the pristine and doped chromium arsenide in various crystal structures to investigate the structural, magnetic, and optical properties for real applications.
By first-principles calculations, we show the structural parameters, the ground magnetic state, density of states, and the optical properties
of CrAs in six different structures. The results regarding structural parameters and magnetic properties agree well with experimental data and other theoretical works.
First, We found that the ground-state structure is an orthorhombic MnP-type structure with antiferromagnetic spin order. The rocksalt structure is an unreported metastable phase and a ferromagnetic metal with high spin polarization at the Fermi level.
Secondly, the NiAs structure and MnP structure have a higher absorption coefficient than other structures in the infrared region and ultraviolet region, respectively. In the visible light region, the wurtzite and zincblende structures are more transparent than other structures.
At last, we found that Ti substitution of Cr and Te substitution of As can lead to a phase transition in ground-state structure and ground-state magnetic order, respectively. These results can promote the application of the CrAs system into spintronics.

\end{abstract}


\maketitle

\section{Introduction}
Spintronics (or magnetoelectronics) \cite{Zutic-rmp-2004}, which adds the spin degree of freedom to the conventional electronic devices, has several advantages like increasing data processing, decreased electric power consumption, and non-volatility \cite{Prinz-science-1998,Prinz-science-1999,Wolf1488}.
In 1983, de Groot \emph{et al} predicted half Heusler alloys NiMnSb and PtMnSb \cite{Groot-prl-1984} to be Half-metallic (HM) materials, which have 100\% spin polarization at Fermi level.
Transition metal pnictides (e.g. CrAs and VAs) have various phases including zincblende (ZB), wurtzite (WZ), hexagonal NiAs (NA), and orthogonal MnP (MP)-type structures \cite{Charifi2018}. These materials have rich magnetic orders and compatibility with the traditional group III-V or II-VI semiconductors(e.g. GaAs, ZnTe)\cite{Charifi2018}. Thus, it is significant to explore the electric and magnetic properties of transition metal pnictides \cite{NA-MP}.

Chromium arsenide (CrAs) exhibit an MP-type orthorhombic structure at low temperature \cite{Selte-1971,Watanabe-jap-1969}.
T. Ito et al predicted that MP-type CrAs shows a double-spiral antiferromagnetic (AFM) magnetic order with the magnetic moment of 1.7 $\mu_{B}$ per Cr atom\cite{Itao-2007}.
Kotegawa et al observed a first-order antiferromagnetic-paramagnetic transition at $265$ K, which is interpreted as an electronic transition between localized and collective states \cite{Boller-1971}. Svitlana Polesya et al proved that nearest-neighbor and next nearest-neighbor Cr-Cr interactions in CrAs are ferromagnetic (FM) and AFM, respectively, which can give the magnetic phase transition temperature is about 270 K \cite{Polesya-2013}, consistent with the experiment value \cite{doi:10.7566/JPSJ.83.093702}.
Wu et al. \cite{Wu-2014} and Kotegawa et al \cite{Kotegawa-prl-2014} independently discovered pressure-induced
unconventional superconductivity in the vicinity of antiferromagnetic (AFM) order in MP-type crystal structure CrAs.
Carmine et al predicted that CrAs may be a weakly correlated material with the Fermi surface exhibiting a two-dimensional behavior, which suggests a close link between the appearance of the superconductivity and the external pressure~\cite{Carmine2017}.
Moreover, W. Wu et al observed that MP-type CrAs show a Fermi-liquid behavior with a $T^{2}$-dependent resistivity at low temperature \cite{Wu-2010}.

Much research effort has also been devoted into other meta-stable phases of CrAs.
N. Kazama et al \cite{Kazama-1971} observed that CrAs occurs a structural phase transition from the MnP structure to NiAs structure at high T = 769 K. Another meta-stable ZB-type CrAs has been fabricated on GaAs(001) substrates by molecular-beam-epitaxy (MBE) \cite{Akinaga-jap-2000,Ofuchi-2003}. The experiment confirmed that ZB-type CrAs have well-pronounced HM behavior with high Curie temperature above 400 K \cite{Ofuchi-2003,Shirai-2003}. Xie et al \cite{Xie-prb-2003} predicted WZ-type CrAs to be HM ferromagnet.
Besides, the magnetic properties of CrAs are very sensitive to substitutions of other elements. Suzuki and Ido reported that substitution of only 7.5\% of phosphorus for arsenic in CrAs yields a collapse of the double-spiral magnetic order \cite{Suzuki-1993}. Meanwhile, the transition temperature of CrAs is very sensitively influenced by other 3d metal substitutions for the Cr \cite{Selte-1975,IDO1997164}. The previous theoretical research paid little attention to the doping effect on the structural properties as well as the optical properties of CrAs system. It is meaningful to explore the structural, magnetic, optical and electronic structural properties of pristine and doped CrAs in various structures.

Here, we focused on a systematical study of CrAs in various crystal structures utilizing the density functional theory. We performed first-principles calculations based for CrAs in six phases including the Rocksalt (RS), Cesium-chloride (CC), ZB, WZ, NA, and MP phase. Then we displayed their optical properties. At last, we report the doping effect on the structural and magnetic properties of CrAs. The computational details are given in the next section. The main results and discussion will be presented in the third section. We shall give our conclusion in the fourth section.

\section{Computational details}

Our density functional theory(DFT) calculations were performed by using the Vienna ab initio simulation package(VASP) \cite{Kresse-1996-54,Kresse-1996-6,Kresse-prb-1994}. This code solves the Kohn-Sham equations within the pseudopotential approximation \cite{Payne-rmp-1992}. The generalized gradient approximation(GGA) to the exchange-correlation potential in form of the Perdew, Burke, and Ernzerhof(PBE) \cite{PBE-prl-1996} functional was used in this work. After convergence tests, an energy cutoff of $650$ eV for the plane-wave expansion was selected.
The K-points, sampling over the irreducible Brillouin zone, were generated by Monkhorst-Pack scheme \cite{Monkhorst-prl-1976}.
We used K-points mesh of $9\times9\times9$ for RS, CC, ZB; $11\times11\times7$ for WZ, NA; and $7\times11\times7$ for MP type, respectively.
The convergence of the results concerning k-point sampling has been carefully checked.
To explore the magnetic properties of CrAs, we have calculated the total energy of NM, FM, and different AFM phases for six crystal structures.
All the structures have been optimized to achieve the minimum energy by accurate relaxation of the atomic positions up to the energy accuracy of $10^{-5}$ eV. The optical properties of solids are described by dielectric functions \cite{Wang2019,Xu2017}.

\begin{table*}[!htb]
\centering
\caption{The lattice types (LT), space group (SG), magnetic state (MS), lattice constant a(\AA), length of Cr-As bond L(\AA), bulk module B (GPa), magnetic moment per formula unit M ($\mu_{B}$), cohesive energy and per CrAs pair E$_{c}$(eV), and relative energy to MP-type structure E$_{m}$ (eV). The abbreviation TW and OW represent the this work and other work, respectively.}
\label{tab}
\begin{tabular*}{0.9\textwidth}{@{\extracolsep{\fill}}ccccccccccccccccccccccccc}
\hline
\hline
Results&  LT                 & SG       & MS         & a/b/c      &  L    &    B    & M   & E$_{c}$  &E$_{m}$  \\
\hline
TW                           & MP &62  &AFM &5.5705/3.4981/6.2702 &2.5562 & 83.33   &  - &-7.418 &0  \\
OW\cite{Selte-1971}          & MP &62  &AFM &5.5773/3.5731/6.1289 &-      & -       &  - &-      &- \\
OW\cite{Podloucky-jmmm-1984} & MP &62  &AFM &5.637/3.445/6.197    &-      & -       &  - &-      &-  \\
OW\cite{Saparov_2012}        & MP &62  & NM &5.649/3.465/6.209    &-      & -       &  - &-      &-  \\
TW                           & NA &194 &AFM &3.7793/5.3234      &2.5558 & 135.36  &  - &-7.377 &0.041 \\
OW\cite{Podloucky-jmmm-1984} & NA &194 &AFM &3.628/5.561        &-      & -       &-      &  -    &-  \\
OW\cite{Motizuki_1986} & NA &194 &NM &3.55/5.68        &-      & -      &-        &  -    &-  \\
TW                           & CC &221 &FM  &3.0003              &2.5983 & 147.62   &1.183  &-6.792 &0.626  \\
TW                           & WZ &186 &FM  &4.15/6.71           &2.5178 & 68.65    &3.000  &-6.533 &0.887  \\
OW\cite{Xie-prb-2003}        & WZ &186 &FM  &4.002/6.530         &-      &68.4      &3.000   & -     &- \\
TW                           & RS &225 &FM  &5.1205              &2.5603 & 94.86    &2.931  &-7.026 &0.392  \\
TW                           & ZB &216 &FM  &5.6683              &2.4545 & 68.83    &3.000  &-6.533 &0.885  \\
OW\cite{Xie-prl-2003}        & ZB &216 &FM  &5.659               &-      & 71.0     &3.000  &-6.533 &- \\
  \hline
  \hline
  \label{table:E}
\end{tabular*}
\end{table*}

\section{Results and Discussions}
\subsection{Structural analysis of six structures}
(i) RS-type CrAs has a space group of $Fm\bar{3}m$ (No.$225$). It composes of two face-centered cubic lattices, shifted by the vector (0.5, 0.5, 0.5), which results in octahedral coordination for both cations and anions.

(ii) C-type CrAs has a space group of $P6_{3}mc$ (No.$221$). The CC structure can be described as two interpenetrating simple cubic lattices shifted by the vector (0.5, 0.5, 0.5) along the body diagonal of the conventional cube. Each anion in a given sublattice has $8$ nearest neighboring cations on the other sublattice, there are $6$ second nearest neighbors of the same type sublattice.

(iii) ZB-type CrAs has a space group $F\bar{4}3m$ (No.$216$). The ZB structure is similar to that of diamond with two interpenetrating face-centered-cubic lattices displaced by the vector (0.25,0.25,0.25) along the body diagonal of the conventional cube. The coordination number is $4$ for each atom.

\begin{figure}[tb]
\centering
\scalebox{0.3}{\includegraphics{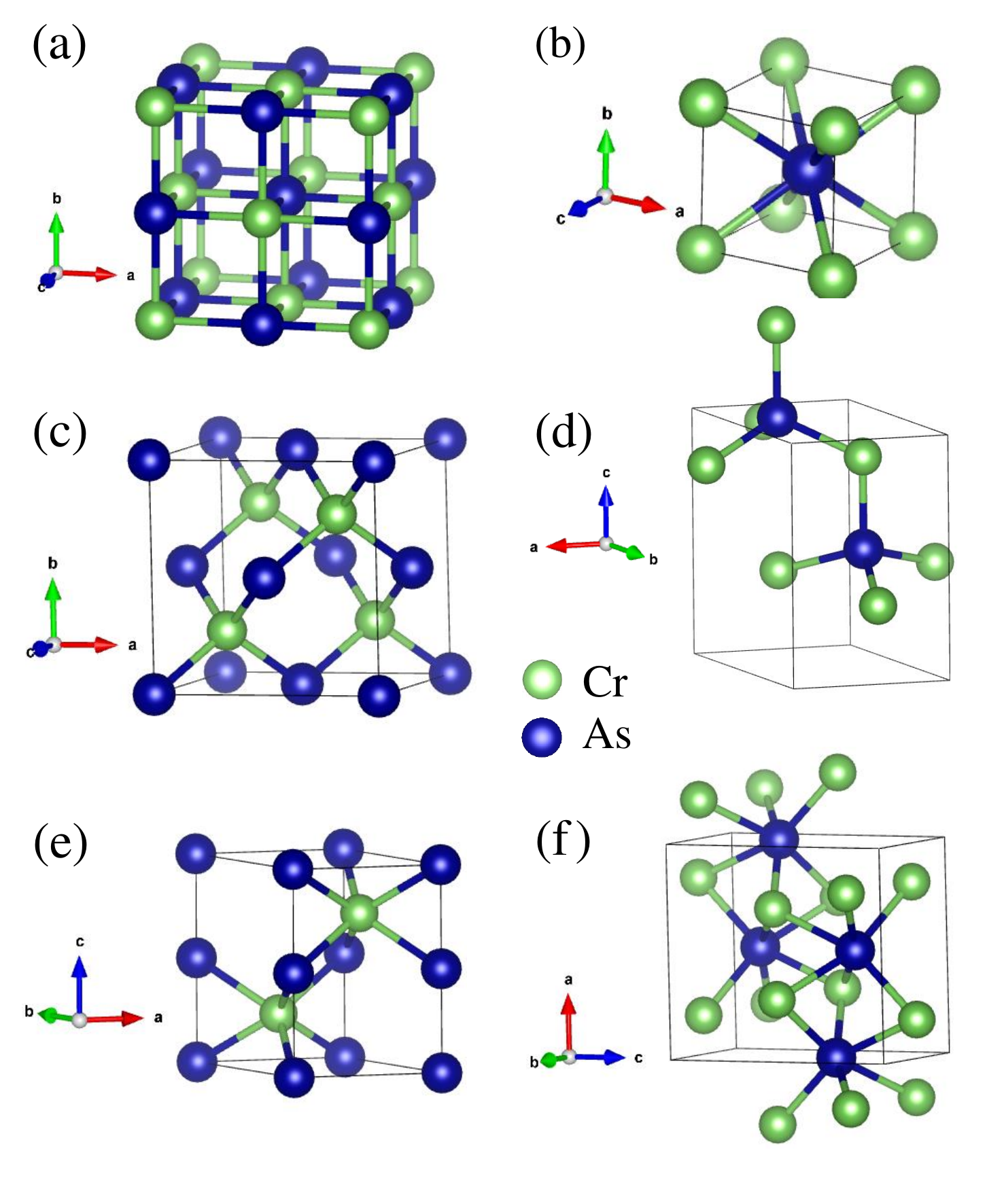}}
\caption{Six types of crystal structures of CrAs. (a) the cubic Rockaslt (RS) structure; (b) the cubic CsCl (CC) structure; (c) the cubic Zinblende (ZB) structure; (d) the hexagonal Wurtzite-type (WZ) structure; (e) the hexagonal NiAs-type (NA) structure; and (f) the orthorhombic MnP-type structure.}
\label{fig:str}
\end{figure}

(iv) WZ-type CrAs has a space group $P6_{3}mc$ (No.$186$). The WZ structure is regarded as a distorted ZB-type structure with small shifts of atomic positions. There are four atoms of two different types in the unit cell (see Figure~\ref{fig:str}(d)). Each atom is surrounded by an tetrahedron of atoms of opposite sort hexagonal analog to the ZB structure and has the same local environment if one assumes the ideal c/a ratio and internal parameters u of 3/8 for the anion site.
\newline
(v) NA-type CrAs has a space group $P6_3/mmc$ (No.$194$), as shown in Figure~\ref{fig:str}(f). It has four atoms in the primitive unit cell. As atoms form a lattice like the hexagonal close-packed structure, and Cr atoms a simple hexagonal lattice. As distortion and shifts are small, the NA and MP structures are similar. However even the slight difference affects magnetic properties through the change of electronic structure.

(vi) MP-type CrAs has a space group $Pnma$ (No.$62$). The orthorhombic MnP-type structure is regarded as distorted NiAs-type structure with small shifts of atomic positions\cite{Motizuki-2009}. The structure, illustrated in Figure~\ref{fig:str}(e), shows that each Cr atom is coordinated to six As atoms in irregular octahedral coordination. Each As atom is surrounded by six Cr atoms, arranged at the corners of a distorted trigonal prism. This structure possesses a space-inversion symmetry.

\subsection{The ground-state and metastable structures}
In the fully optimized calculations, all lattice parameters and internal atomic positions were relaxed to find out the ground state of CrAs. By fitting the Birch-Murnaghan equation of state to the obtained total energies as a function of volume, the structural properties of the systems were determined.

\begin{figure}[bht!]
\centering
\scalebox{0.25}{\includegraphics{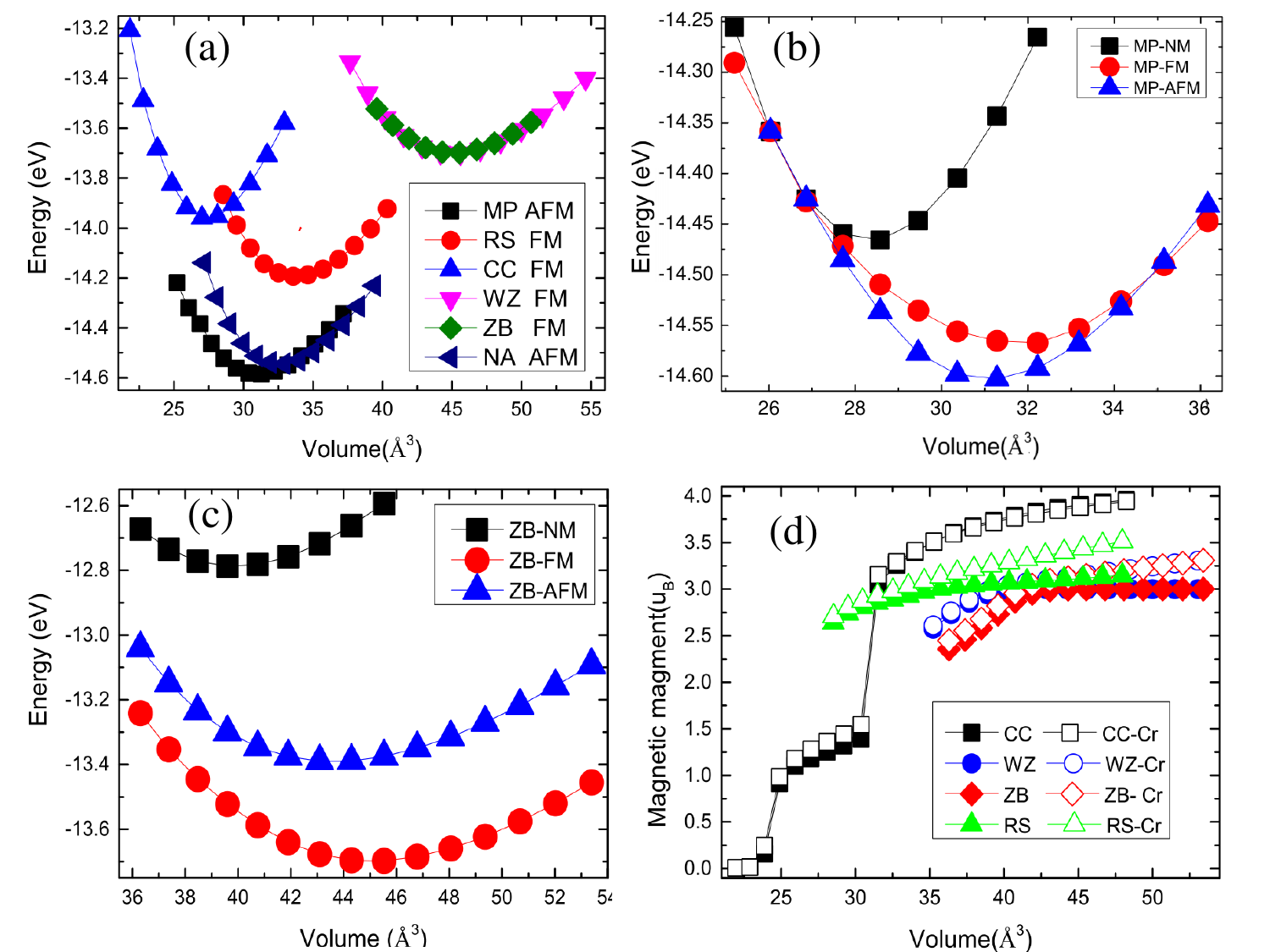}}
\caption{(a) Total energy as a function of volume for CrAs in six crystal structures. For every structure only the lowest phase of NM, AFM and FM is shown. Energy as a function of volume of CrAs in the NM, FM and AFM states at (b) MP-type and (c) ZB-type structure, respectively. (d) Magnetic moment-volume curves of four crystal structure CrAs in FM state.}
\label{fig:ev}
\end{figure}

We first investigated the NM, FM, and AFM states of CrAs in the MP-type structure. As seen in Figure~\ref{fig:ev}(b), the MP-type CrAs in AFM state is energetically more stable compared to the FM and NM state over a wide range of lattice volumes. This is consistent with the experimental observation that MP-type CrAs is in AFM state\cite{Hashemifar-prb-2010}. The equilibrium lattice volume of AFM MP-type CrAs is found to be 31.28 \AA$^3$.

Then the calculated total energies versus formula volume for CrAs in six crystal structures are displayed in Figure~\ref{fig:ev}(a). For the MP and NA structures only AFM phases are presented because the NM and FM phases are always higher in total energy (see Figure~\ref{fig:ev}(b)).
For the other four structures: RS, CC ZB and WZ, we choose ZB as a representative. The calculated total energies as a function of crystal volume for NM, FM,and AFM states of CrAs in ZB-type structure are plotted in Figure~\ref{fig:ev}(c). As seen in this figure, the total calculation indicates that the FM state is most stable for this structure. This is consistent with the previous work that ZB-type CrAs is an FM state\cite{Xie-prl-2003,Akinaga-jap-2000}
Thus, only the FM phases of ZB, CC, WZ, RS are presented in Figure~\ref{fig:ev}(a) because the NM and AFM phases are always higher in total energy. The results of structural properties and magnetic moments of six structures for CrAs together with theoretical and experimental results are summarized in Table~\ref{table:E}.

As seen from Table~\ref{table:E}, it can be seen that the fully optimized MP-type CrAs in AFM state is lower in energy than any other phases, so it is the ground-state phase, this agrees well with the experimental ground state, and theoretical structural parameters in this state are in agreement with experimental values\cite{Selte-1971}.
The NA-type CrAs in AFM state is $40$ meV higher than MP-type CrAs in the AFM state (see Figure~\ref{fig:ev}(a)).
However, the MP-type CrAs in the FM state has a higher energy of $36$ meV than MP-type CrAs in the AFM state (see Figure~\ref{fig:ev}(b)).
Thus, the energy of MP-type CrAs in the FM state is slightly smaller than the NA-type CrAs in AFM state and is the second-most stable phase.
This may explain that the new metastable FM orthorhombic structure CrAs epilayers can be grown on GaAs(001) in a recent experiment\cite{Etgens-prl-2004}, and it also agrees with the results of Hashemifar's \cite{Hashemifar-prb-2010}.

The metastable structure with the energy next to that of MP and NA structure is the CC structure (see Figure~\ref{fig:ev}(a)) which prefers an FM state.
The ZB-type CrAs, which has been synthesized in the experiment\cite{Ofuchi-2003}, is 0.885 eV higher in energy than MP-type AFM state, it agrees well with former's work\cite{Hashemifar-prb-2010}. The equilibrium energy and volumes of the FM ZB-type and FM WZ-type are almost the same, with their difference being only 2 meV and 0.09{\AA}$^{3}$, respectively.

\subsection{The magnetic properties of CrAs}
CrAs in ZB, CC, WZ and, RS phase prefer the FM state. We calculated the magnetic moments of those four structures in the FM phase at different volumes, and the results are presented in Table~\ref{table:E} and Figure~\ref{fig:ev}(d).
The magnetic moment for ZB-type and WZ-type CrAs in FM state was calculated as 3 $\mu_{B}$ per CrAs pair at the equilibrium volumes, which is in excellent agreement with the saturation moments estimated experimentally\cite{Akinaga-jap-2000} and theoretical results\cite{Xie-prl-2003}.
The net moment of 3.0 $\mu_{B}$ results from the remaining four of the 3d plus two of 4s electrons of chromium bonding with the three As p electrons. The magnetic moments of ZB-type and WZ-type CrAs remain unchanged under compressive strain down to -12\% and -5\% in relative volumes, respectively.

RS-type CrAs has a magnetic moment of $2.9 \mu_{B}$ and $3 \mu_{B}$ per CrAs pair at the equilibrium volumes and expanded volumes, respectively.
CC-type CrAs has a magnetic moment of only 1.183 $\mu_{B}$ per CrAs pair at the equilibrium volumes. However, the magnetic moments of CC-FM grow rapidly with increasing volume.

\begin{figure}[tb]
\centering
\scalebox{0.50}{\includegraphics{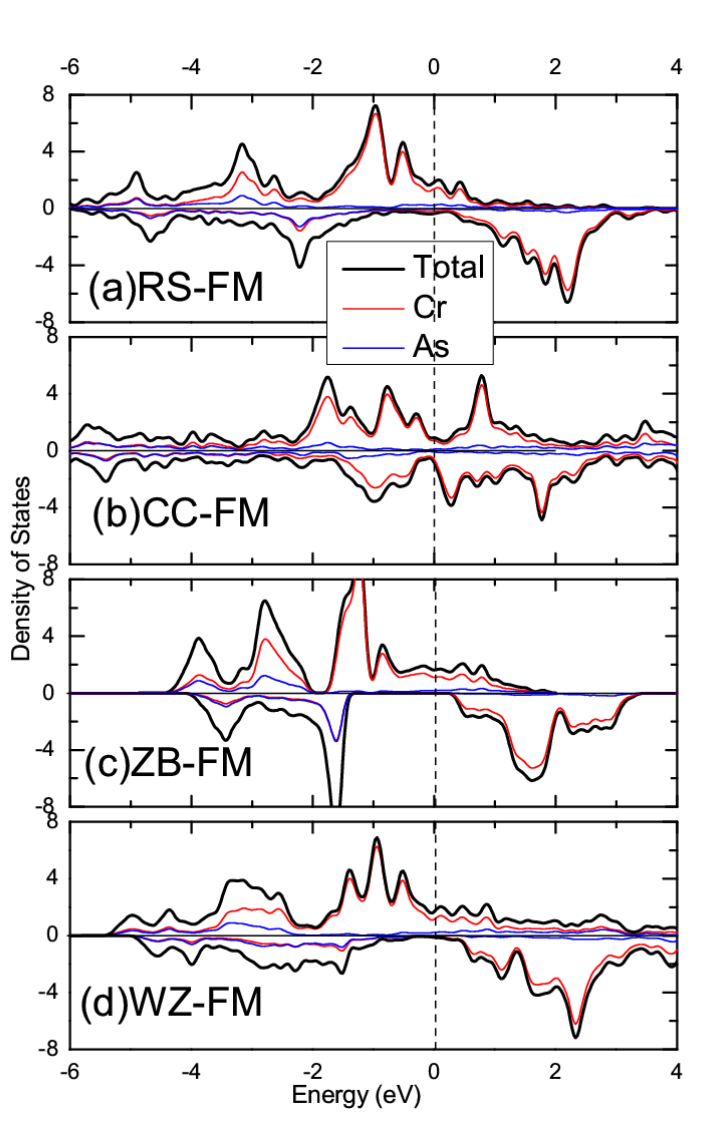}}
\caption{Spin-dependent total densities of states (DOS, states/eV per formula unit) of the four FM CrAs in stable magnetic state. Upper and lower parts of every panel are for majority-spin and minority-spin, respectively.}
\label{fig:dos}
\end{figure}

\subsection{Electronic properties of CrAs}
To study the electronic structures of the system, the density of states (DOS) are calculated with these optimized structures. Spin-resolved DOS of in RS-, CC-, ZB- and WZ-type CrAs in the FM state are presented in Figure~\ref{fig:dos}.
The Cr-d orbitals dominated the DOS around the Fermi level.

We defined the E$_{bc}$ and E$_{tv}$ as the bottom energy of minority-spin conduction bands and absolute values of the top energy of minority-spin valance bands with respect to the Fermi energy, respectively. The half-metallic (HM) gap is defined as the minimum of E$_{bc}$ and E$_{tv}$.
In ZB and WZ phases, there are clear non-zero HM gap, which is essential to a half-metallic ferromagnet with 100\% spin-polarized carriers.
Large HM gaps in WZ-type FM phase reaches $0.49$ eV, which agrees well with former's work\cite{Xie-prb-2003,Xie-prl-2003}.
For the mechanism of HM in the ZB and WZ structures of CrAs, please see the Ref\cite{Xie-prb-2003,Xie-prl-2003,Mavropoulos_2007}.

Figure~\ref{fig:dos}(c) displayed the DOS of RS-type CrAs. The DOS at minority-spin channel is non zero. The lack of HM gap around the Fermi level indicate that RS-type structure is not HM, but has a high spin polarization. Contrast with RS, ZB, and WZ structures, CC-type CrAs has s small a spin polarization of $2.8\% $ at the Fermi level.

\begin{figure}[tb]
\centering
\scalebox{0.3}{\includegraphics{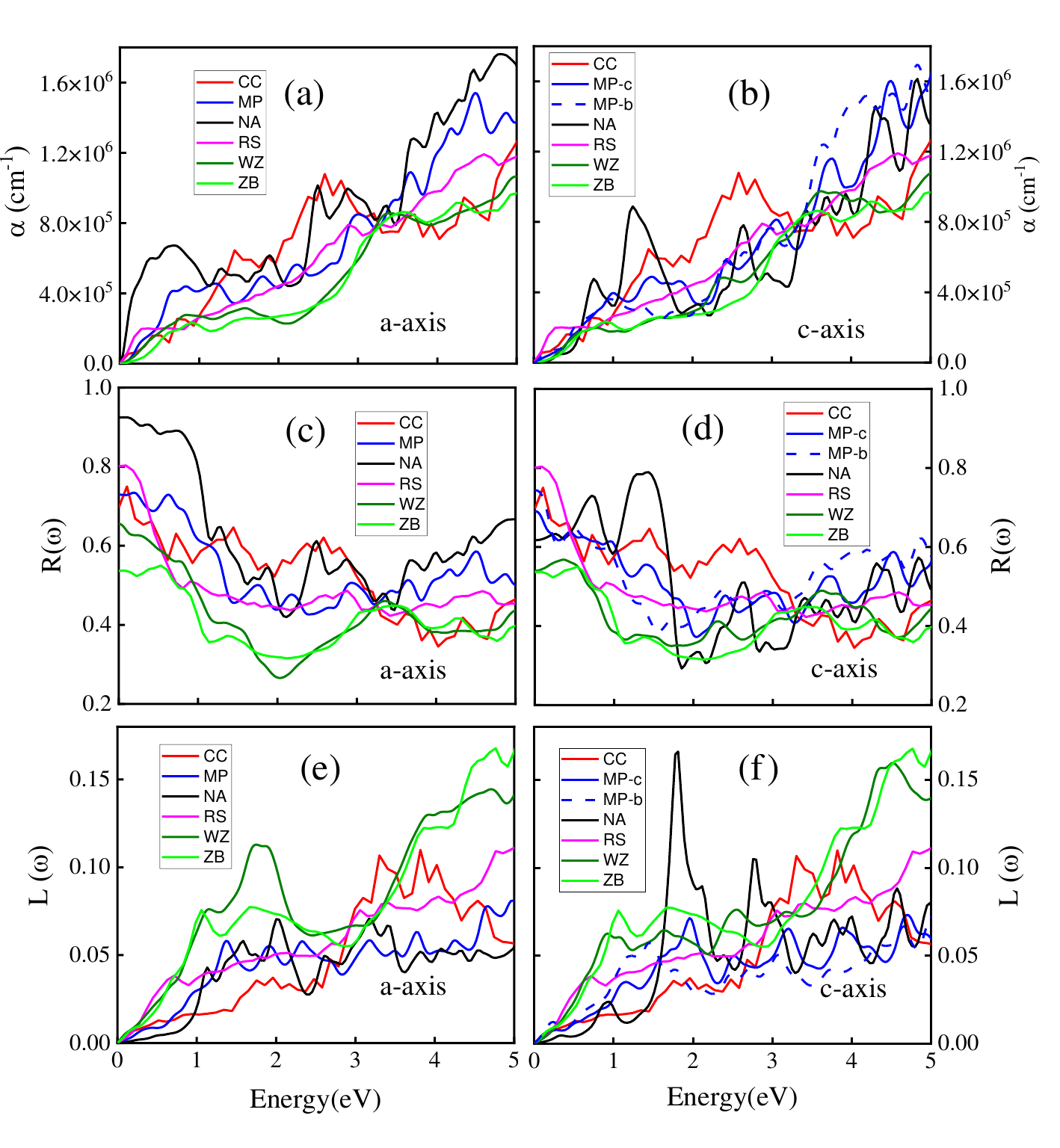}}
\caption{Spectra of different phases of CrAs in the direction of a-axis and c-axis (including b-axis): (a) and (b) absorption coefficient; (c) and (d) reflectivity; (e) and (f) energy loss spectrum.}
\label{fig:optical}
\end{figure}

\subsection{Optical properties of CrAs}

We further calculated the dielectric function to study the optical properties of CrAs.
Since the CC-, ZB- and RS- type CrAs have isotropic lattices, only the results for the a-axis are presented.
Considering the anisotropy lattice of MP, NA, and WZ structure, the results for the a-axis and c-axis (including b-axis for MP) are displayed.

As shown in Figure~\ref{fig:optical}(a) and \ref{fig:optical}(b), the absorption coefficient $\alpha$ of NA-type CrAs in the infrared region (0.6 eV-1.3 eV) is higher than that of other phases along different axis. In the ultraviolet region (3.5 eV-5 eV), the NA structure has a higher $\alpha$ along the a-axis, and the MP structure has a slightly higher $\alpha$ along the b-axis.

Based on Figure~\ref{fig:optical}(c) and ~\ref{fig:optical}(d), it can be seen that NA structure has a higher reflectivity $R$ in the infrared region (0.6 eV-1.3 eV). The $R$ of NA structure along the c axis is $78.9\%$ at 1.435 eV, which is significantly higher than other phases. In the visible light region (1.64 eV-3.19 eV), the $R$ of WZ and ZB  structures along the a-axis is lower than other phases, and there is no such phenomenon on the c-axis. And the $R$ of NA and MP structures along the a-axis in the ultraviolet region (3.5 eV-5 eV) is significantly higher than other phases.

The energy loss spectrum $L$ corresponds to the reflection coefficient $R$, and the position with the largest energy loss corresponds to the position where the reflection coefficient drops sharply. At about 1.8 eV, the reflection curvature of NA structure along the c axis decreases, and the energy loss coefficient has a peak at the same energy position. Similarly, the reflection curve of CC and WZ structures drops about 3.0 eV and 1.8 eV along the a-axis, respectively. As a results, the energy loss coefficient has a peak at the same energy position.

\subsection{Doping effect}
At last, we investigated the effect of external doping on the structural and magnetic properties of CrAs.

\begin{figure}[tb]
\centering
\scalebox{0.3}{\includegraphics{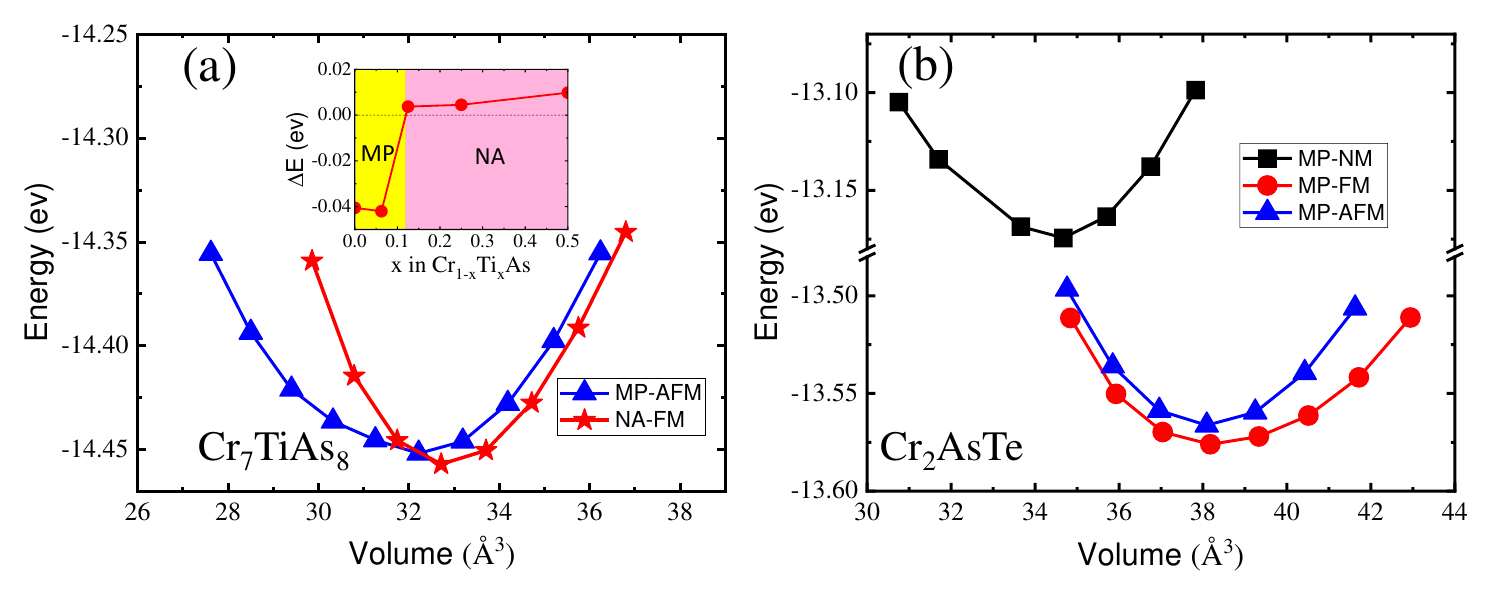}}
\caption{(a) Total energy as a function of volume for Cr$_{7}$TiAs$_{8}$ in MP-type and NA-type structure. For every structure only the lowest phase of NM, AFM and FM is shown. The inset presents the difference between the MP-type and NA-type $\Delta$E=E$_{\rm MP}$-E$_{\rm NA}$ for the Cr$_{\rm 1-x}$Ti$_{\rm x}$As system. (b)Total energy as a function of volume for Cr$_{2}$AsTe in NM, FM, AFM state at MP-type structure. }
\label{fig:doped}
\end{figure}

Firstly, we substituted the Cr atom in CrAs by Ti stom at different doping concentration $x$. We calculated the energy of the compound Cr$_{\rm 1-x}$Ti$_{\rm x}$ in different structures and different magnetic state. We found that MP and NA structures still have lower energy than other structures. The phase diagram of the CrAs system is shown as the inset of Figure~\ref{fig:doped} (a), which is drawn based on the energy difference between the MP-type and the NA-type $\Delta E = E_{\rm MP}-E_{\rm NA}$. It is found that the ground-state crystal structure is the orthorhombic MP-type when the doing concentration of $x \leqslant 0.12$. However, the compound Cr$_{\rm 1-x}$Ti$_{\rm x}$ prefers the hexagonal NA-type structure for x$>$0.12. We have examined the MP-type and NA-type Cr$_{\rm 1-x}$Ti$_{\rm x}$ compound in different magnetic state. The energy-volume curve of $x = 0.125$ at which CrAs exhibit a NA structure. It is shown that the ground-state phase is FM state. The electronic state calculation indicates that Ti-doped CrAs at $x = 0.125$ is a magnetic metal.

Then we considering the substitution of the As atom by Te atom at different doping concentration $x$. We found that the compound CrAs$_{1-x}$Te$_x$ still maintained the orthogonal MnP-type structure as ground-state structure but has a magnetic phase transition from AFM to FM at x=0.5, as shown in Figure~\ref{fig:doped} (b).  The electronic state calculation indicates that Te-doped CrAs at $x = 0.5$ is also a magnetic metal. We noticed that doping concentration in CrAs have reach more than 50 \%~\cite{IDO1997164}, so the structural and magnetic phase transition might be observed in future experiment and be applied in the spintronic devices.

\section{Conclusion}
We adopted the first-principles calculations to study
the structural, magnetic, optical properties, and doping
effect in CrAs system with different structures. It was found that the FM state is energetically favorable for RS-type, CC-type, ZB-type, and WZ-type structure, while AFM state is energetically favorable for NA-type and MP-type structure. MP-type structure in AFM state is the ground-state structure for CrAs.
%dos
The density of states show that ZB-type and WZ-type structures exhibit HM behavior with a band gap in spin-down channel. RS-type and CC-type CrAs is not HM and has a large and small spin-polarization at the Fermi level, respectively.
%optics
The optical properties show that the NA-type has a higher absorption coefficient and reflectivity in the infrared region at all the axis. NA-type and MP-type have high absorption coefficients and reflectivity in the ultraviolet region along the a-axis.
WZ-type and ZB-type have small absorption coefficients in the visible light region.
%doped
At last, we found that Ti-doped CrAs occur a structural phase transition from MP-type to NA-type structure at a doping concentration of about $x=0.12$. The Te-doped CrAs can maintain the structure and exhibit an AFM-to-FM phase transition at a doping concentration of about $x=0.5$.
We hope our results can stimulate further interest in both theoretical and experimental research in the binary compound CrAs.

\begin{acknowledgments}
This work was supported by National Natural Science Foundation of China (No. 11904312 and 11904313),
the Project of Department of Education of Hebei Province, China(No. BJ2020015), and the Natural Science Foundation of Hebei Province (No. A2019203507 and A2020203027).
K.C. Zhang acknowledges the fund support from LiaoNing Revitalization Talents Program (No. XLYC2007120). The authors thank the High Performance Computing Center of Yanshan University.
\end{acknowledgments}

\nocite{*}
\bibliography{cras}

\end{document}